\documentclass[aps,
preprint,
longbibliography]{revtex4-2}
\usepackage[pdftex]{graphicx}
\usepackage{amsmath,amssymb}
\usepackage{bm}% bold math
\usepackage{tcolorbox}
\usepackage[utf8]{inputenc}
\usepackage{etoolbox}
\usepackage{siunitx,physics,float,mathtools}
\usepackage{empheq}
\usepackage{here,overpic}
\usepackage{xcolor}
\usepackage{natbib}
\usepackage{url}
\usepackage{textcase}
\usepackage{oplotsymbl}
\begin{document}
% \preprint{APS/123-QED}

%\preprint{}

\title{Relationship between the power spectral density of the Lagrangian velocity and the hierarchy of coherent vortices in turbulence}
% \title{Lagrangian velocity spectrum in terms of the hierarchy of coherent vortices in turbulence}


\author{Yusuke Koide}
\email{koide.yusuke.k1@f.mail.nagoya-u.ac.jp}
\affiliation{ 
Graduate School of Engineering, Nagoya University, Furo-cho, Chikusa, Nagoya, Aichi 464-8603, Japan%\\This line break forced with \textbackslash\textbackslash
}%
%  \altaffiliation[Also at ]{Physics Department, XYZ University.}%Lines break automatically or can be forced with \\
\author{Susumu Goto}%
%\thanks{}
\affiliation{Graduate School of Engineering Science, Osaka University, 1-3 Machikaneyama, Toyonaka, Osaka 560-8531, Japan}


\date{\today}

\begin{abstract}

We conduct direct numerical simulations of developed turbulence in a periodic cube to investigate the formation mechanism of the power spectral density of the Lagrangian velocity.
We compare the power spectral density of the Lagrangian velocity of turbulent flows with different forcing methods and Reynolds numbers.
This systematic comparison demonstrates that universal behavior is observed in a narrow high-frequency regime, whereas non-universality originating from the forcing method broadly appears in a low-frequency regime.
To reveal the formation mechanism of the spectra in terms of the hierarchy of coherent structures in turbulence, we propose a scale-decomposition method for the Lagrangian velocity, which enables us to evaluate the contribution of vortices at different scales.
This scale-decomposition analysis directly demonstrates that the largest-scale flows driven by the external force can contaminate the Kolmogorov scaling of the Lagrangian velocity spectra formed by small-scale vortices in the inertial range, thus leading to the narrow Lagrangian inertial range.
Furthermore, we provide evidence that this remarkable effect by the largest-scale flows is specific to the Lagrange velocity by demonstrating that the power spectral density of the Eulerian velocity is less sensitive to the forcing method.

\end{abstract}

%\keywords{}

\maketitle
\newpage

\section{\label{sec:Intro}Introduction}

The Lagrangian viewpoint provides valuable insight into the physics of turbulence.
From a theoretical perspective, the closure approximation using the Lagrangian timescale instead of the Eulerian timescale successfully derives a Kolmogorov scaling $E(k)\propto k^{-5/3}$ of the energy spectrum $E(k)$~\cite{Kraichnan1965-nd,Kaneda1981-ei,Kida1997-fj}.
In addition, the Lagrangian approach is essential for particle dispersion and mixing in turbulence~\cite{Yeung2002-xy,Toschi2009-ua}.
Another interesting example is turbulence modulation by a small amount of polymers~\cite{Gyr1995-nn,White2008-gb,Xi2019-nc}.
Since the dynamics of polymers in turbulence is dominated by the Lagrangian history~\cite{Watanabe2010-oi}, detailed knowledge of the Lagrangian statistics is required for understanding the physical mechanism of turbulence modulation by polymers.

Although the Eulerian spatial statistics of turbulence at high Reynolds numbers preceded the Lagrangian statistics, owing to the development of experimental techniques and computational power, there has been an increasing amount of literature on the Lagrangian velocity over the past decades~\cite{Yeung2002-xy}.
One of the key issues is the Kolmogorov scaling in the Lagrangian velocity.
The Kolmogorov similarity hypothesis predicts that the power spectral density $E_L(\omega)$ of the Lagrangian velocity follows $E_L(\omega)\propto \omega^{-2}$ in the Lagrangian inertial range.
Previous studies found this scaling law using experiments and direct numerical simulations~(DNS)~\cite{Hanna1981-th,Lien1998-ij,Mordant2001-td,Yeung2001-rq,Mordant2004-uk,Chevillard2005-qe,Yeung2006-nl,Ouellette2006-gd,Huang2013-nv,Angriman2020-gq}.
Similarly, many studies have attempted to confirm the Kolmogorov scaling in the second-order Lagrangian structure function $S_2(\tau)=\langle |u(t+\tau)-u(t)|^2\rangle$.
However, since a very high Reynolds number is necessary to yield a clear scaling law $S_2(\tau)\propto \tau$~\cite{Lien2002-bm}, most studies reported the compensated structure function $S_2(\tau)/\tau$ which only shows a peak instead of a plateau region~\cite{Mordant2004-uk,Yeung2006-nl,Ouellette2006-gd,Sawford2011-jp,Falkovich2012-nj,Huang2013-nv,Lanotte2013-ax,Angriman2020-gq}.
An ambiguous scaling law in $S_2(\tau)$ is attributed to the contamination from large-scale flows~\cite{Huang2013-nv,Angriman2020-gq}.
In fact, Angriman \textit{et al.}~\cite{Angriman2022-hs} confirmed a clear power-law behavior even at moderate Reynolds numbers by introducing the multitime structure function, which removes the slow fluctuations due to large-scale flows.

From the Eulerian viewpoint, coherent structures of vortices with different length scales are one of the essential aspects of turbulence.
Thus, the relationship between the Eulerian spatial characteristics and the Lagrangian dynamical characteristics has received considerable attention~\cite{Corrsin1959-on,Kamps2009-gw,Rosso2014-za,He2017-al}.
Lucci \textit{et al.}~\cite{Rosso2014-za} derived an approximate relation between $E_L(\omega)$ and $E(k)$ using the Belinicher--L'vov coordinate~\cite{belinicher1987scale} to remove the sweeping effect.
Although they impose some assumptions to obtain an explicit relation, their method successfully reproduced $E_L(\omega)$ from $E(k)$.
On the other hand, several studies focused on the correlation function of the Fourier coefficient of the Lagrangian velocity.
Previous studies~\cite{Gotoh1993-qm,Kaneda1999-rl,Matsumoto2021-dg} found that the correlation function at wave number $k$ decayed with a characteristic timescale $\tau_L^{(k)}\propto k^{-2/3}$, which is consistent with the Lagrangian timescale derived from the Kolmogorov similarity hypothesis.
Nevertheless, it is still unknown how the power spectral density $E_L(\omega)$ of the Lagrangian velocity is formed by vortices with different length scales in turbulence.

This study aims to directly reveal the relationship between $E_L(\omega)$ and the hierarchy of coherent vortices in turbulence.
For this purpose, we develop a scale-decomposition method for the Lagrangian velocity.
A key point of our approach is applying the Fourier bandpass filter to the Eulerian velocity field instead of decomposing the Lagrangian time series itself.
The present method allows us to evaluate the effect of different-scale vortices on the dynamics of the Lagrangian velocity.
First, we conduct DNS of turbulent flows with various forcing methods and Reynolds numbers to clarify the universality and non-universality of $E_L(\omega)$.
Then, we demonstrate the underlying mechanism of the universality and non-universality of $E_L(\omega)$ by decomposing $E_L(\omega)$ into the contributions from different scales.
Furthermore, we reveal the properties specific to the Lagrange velocity through a systematic comparison with the energy spectra and the power spectral density of the Eulerian velocity probed at a fixed position.
\section{\label{sec:Method}Direct numerical simulation}

To investigate the Lagrangian velocity in fully developed turbulence, we perform DNS of an incompressible Newtonian fluid under periodic boundary conditions in three orthogonal directions with period $2\pi$.
The three-dimensional Navier--Stokes equation with an external force $\bm{f}(\bm{x},t)$ is numerically solved using the Fourier spectral method.
We use three types of external force to examine the universality and non-universality of the Lagrangian velocity spectra.
The first is the steady force $\bm{f}^\mathrm{(V)}(\bm{x})$ expressed as
\begin{equation}
    \bm{f}^\mathrm{(V)}(\bm{x}) = (-\sin x\cos y,\cos x\sin y,0)^\mathsf{T}.
\end{equation}
The forcing wave number $k_f$ of $\bm{f}^\mathrm{(V)}$ is $\sqrt{2}$.
The second is the time-dependent force $\bm{f}^\mathrm{(I)}(\bm{x},t)$ whose Fourier coefficient $\widehat{\bm{f}}^{(\mathrm{I})}(\bm{k},t)$ is expressed as 
\begin{equation}  \label{eq: cases f}
    \widehat{\bm{f}}^{(\mathrm{I})}(\bm{k},t)=
        \begin{dcases}
            \frac{P}{2E_{k_f}(t)}\widehat{\bm{u}}(\bm{k},t)    &   0<|\bm{k}|\leq k_f  \\
            0        &   \text{otherwise},
        \end{dcases}
\end{equation}
where $P$ is the energy input rate, $\widehat{\bm{u}}(\bm{k},t)$ is the Fourier coefficient of $\bm{u}(\bm{x},t)$, and $E_{k_f}(t)$ is defined as 
\begin{equation}
    E_{k_f}(t) = \sum_{0<|\bm{k}|\leq k_f}\frac{1}{2}|\widehat{\bm{u}}(\bm{k},t)|^2.
\end{equation}
The forcing by $\bm{f}^\mathrm{(I)}$ leads to a constant energy input rate $P$~\cite{Lamorgese2005-mf}.
We set $P=1$ and $k_f=2.5$.
The third is the random force $\bm{f}^\mathrm{(R)}(\bm{x},t)$~\cite{Alvelius1999-cr}.
Note that $\bm{f}^\mathrm{(R)}(\bm{x},t)$ is defined as a solenoidal white noise process.
We impose $\bm{f}^\mathrm{(R)}(\bm{x},t)$ on the velocity field in the low-wave-number range centered at $k_f=2$.
The time integration uses the fourth-order Runge--Kutta--Gill scheme, and the phase shift method removes the aliasing errors.

To obtain the Lagrangian data, we track $32^3$ fluid particles uniformly dispersed in the flow.
The position $\bm{x}_L(t|\bm{x}_0,t_0)$ of the fluid particle follows
\begin{equation}
    \frac{d}{dt}\bm{x}_L(t|\bm{x}_0,t_0) = \bm{v}_L(t|\bm{x}_0,t_0), \label{eq:Lagrangian_particle}
\end{equation}
where $\bm{v}_L(t|\bm{x}_0,t_0)$ is the Lagrangian velocity, $t_0$ is the labeling time, and $\bm{x}_0$ is the position of the particle at $t=t_0$. 
Using the Eulerian velocity $\bm{u}(\bm{x},t)$, $\bm{v}_L(t|\bm{x}_0,t_0)$ is expressed as
\begin{equation}
    \bm{v}_L(t|\bm{x}_0,t_0) = \bm{u}(\bm{x}_L(t|\bm{x}_0,t_0),t).
\end{equation}
To obtain $\bm{v}_L(t|\bm{x}_0,t_0)$, the trilinear interpolation is used for the fluid velocity $\bm{u}(\bm{x}_L(t|\bm{x}_0,t_0),t)$.
We integrate Eq.~\eqref{eq:Lagrangian_particle} with the fourth-order Runge--Kutta--Gill scheme.

\begin{figure}
    \centering
    \begin{overpic}[width=0.5\linewidth]{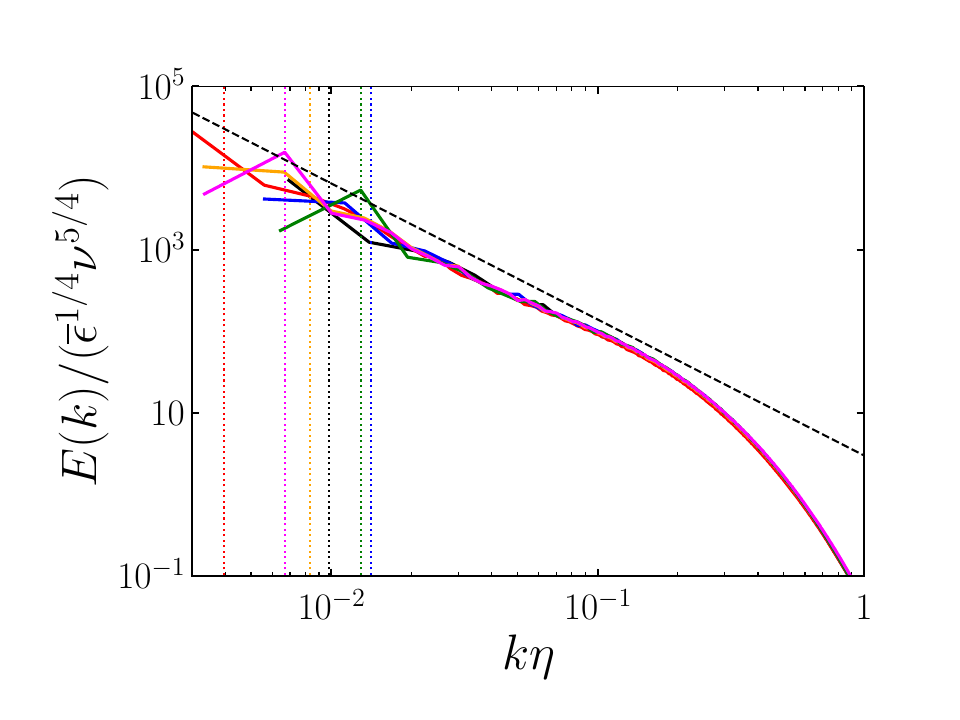} 
    \end{overpic}
    \caption{Energy spectrum $E(k)$ of turbulent flows driven by $\bm{f}^\mathrm{(V)}$ with $\mathrm{Re}_\lambda=117$~(black) and $247$~(red), $\bm{f}^\mathrm{(I)}$ with $\mathrm{Re}_\lambda=217$~(blue) and $312$~(orange), and $\bm{f}^\mathrm{(R)}$ with $\mathrm{Re}_\lambda=205$~(green) and $323$~(magenta). The black dashed line indicates $E(k)\propto k^{-5/3}$. The dotted lines indicate $k\eta=k_f\eta$.}
    \label{fig:ene_spe}
\end{figure}

Table~\ref{table:Parameter} shows the DNS parameters and statistics of turbulence.
In the table, $N^3$ is the number of Fourier modes, $\mathrm{Re}_\lambda$ is the Reynolds number based on the Taylor microscale $\lambda$, $k_\mathrm{max}=\sqrt{2}N/3$ is the largest resolved wave number, and $\eta=(\nu^{3}/\overline{\epsilon})^{1/4}$ is the Kolmogorov length, where $\overline{\epsilon}$ is the mean turbulent energy dissipation rate per unit mass and $\nu$ is the kinematic viscosity.
Here, we define $\mathrm{Re}_\lambda$ as
\begin{equation}
    \mathrm{Re}_\lambda=\sqrt{\frac{20}{3\nu\overline{\epsilon}}}K^\prime,
\end{equation}
where $K^\prime$ is the turbulent kinetic energy per unit mass.
The Courant--Friedrichs--Lewy~(CFL) number is defined as $\sqrt{2K/3}\Delta t/\Delta x$, where $K$ is the kinetic energy per unit mass, $\Delta t$ is the time step, and $\Delta x$ is the grid width.
Figure~\ref{fig:ene_spe} shows the energy spectra $E(k)$ of turbulent flows with the parameters in Table~\ref{table:Parameter}.
Here, $k$ and $E(k)$ are normalized by $\overline{\epsilon}$ and $\nu$.
For $k_f\eta\lesssim k\eta\lesssim 0.1$, $E(k)$ exhibits a power law that is consistent with the Kolmogorov scaling, and the inertial range extends with $\mathrm{Re}_\lambda$.
Although $E(k)$ has a different shape in the low-wave-number regime depending on the forcing method, the effect only appears at $k\lesssim k_f$.

% --------------------
\begin{table}
    \caption{Parameters and statistics of turbulence: $\bm{f}$ is the external force, $N^3$ is the number of Fourier modes, $\mathrm{Re}_\lambda$ is the Reynolds number based on the Taylor microscale $\lambda$, $k_\mathrm{max}=\sqrt{2}N/3$ is the largest resolved wave number, and $\eta$ is the Kolmogorov length.
    The Courant--Friedrichs--Lewy~(CFL) number is defined as $\sqrt{2K/3}\Delta t/\Delta x$, where $K$ is the kinetic energy per unit mass, $\Delta t$ is the time step, and $\Delta x$ is the grid width. }
    \label{table:Parameter}  
    \begin{ruledtabular}
    \begin{tabular}{ccccc}
    $\bm{f}$ & $N^3$&$\mathrm{Re}_\lambda$&$k_\mathrm{max}\eta$&CFL number\\
    \hline

    $\bm{f}^\mathrm{(V)} $&$512^3$ &$117$  & $1.7$&$7.0\times 10^{-2}$ \\
    $\bm{f}^\mathrm{(V)} $&$1024^3$ &$247$  & $1.4$&$7.1\times 10^{-2}$ \\
    $\bm{f}^\mathrm{(I)} $&$512^3$ &$217$  & $1.4$&$5.4\times 10^{-2}$ \\
    $\bm{f}^\mathrm{(I)} $&$1024^3$ &$312$ & $1.6$&$5.5\times 10^{-2}$ \\
    $\bm{f}^\mathrm{(R)} $&$512^3$ &$205$  & $1.6$&$6.6\times 10^{-2}$ \\
    $\bm{f}^\mathrm{(R)} $&$1024^3$ &$323$  & $1.6$&$6.7\times 10^{-2}$ \\
    \end{tabular}
    \end{ruledtabular}
    \end{table}
% --------------------

%   --------------------
%   --------------------
\section{\label{sec:Result}Results}
\subsection{\label{subsec:power_spe}Power spectral density of the Lagrangian velocity}

\begin{figure*}
    \centering
        \begin{tabular}{c}
        \begin{minipage}{0.5\hsize}
            \begin{overpic}[width=1\linewidth]{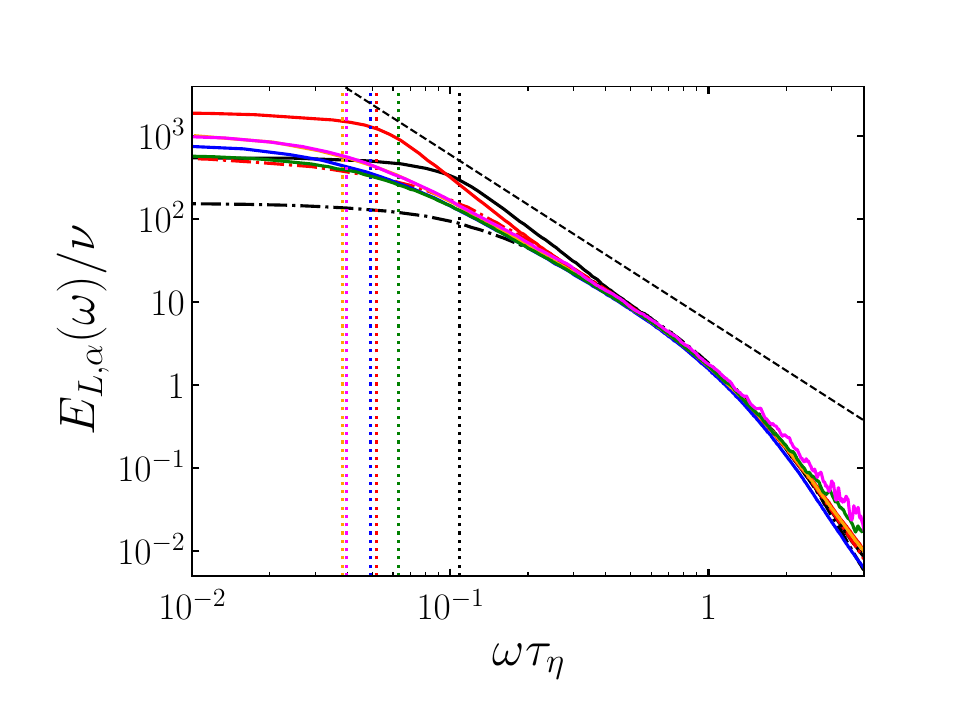}
                \linethickness{3pt}
          \put(5,63){(a)}

            \end{overpic}
        \end{minipage}
        \begin{minipage}{0.5\hsize}
            \begin{overpic}[width=1\linewidth]{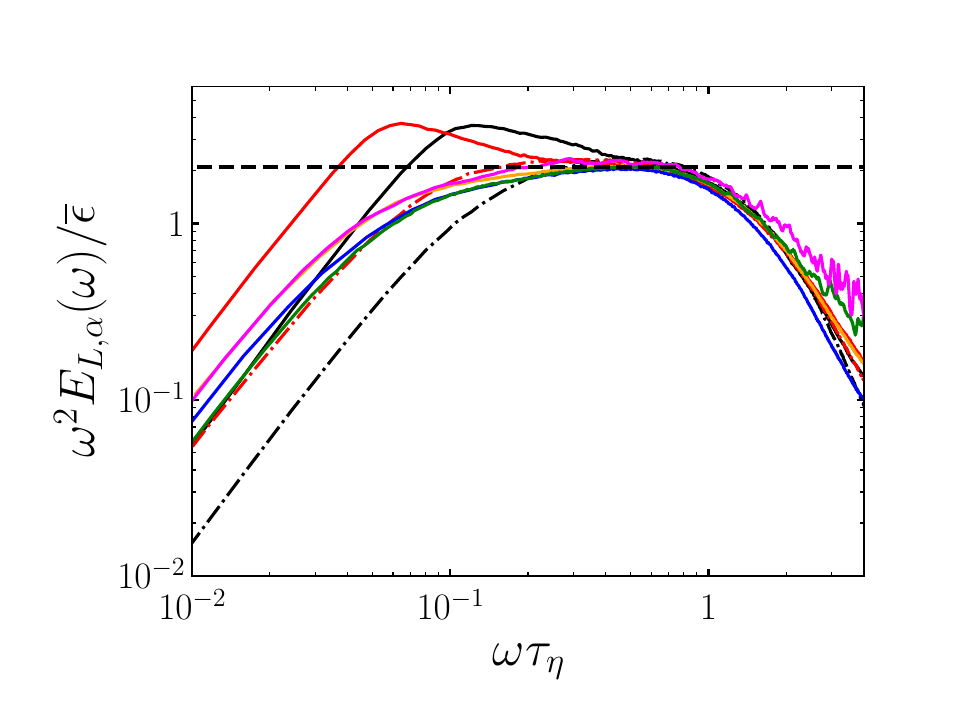}
                \linethickness{3pt}
          \put(5,63){(b)}

            \end{overpic}
        \end{minipage}
        \end{tabular}
    \caption{(a) Power spectral density $E_{L,\alpha}(\omega)$ of the Lagrangian velocity $v_{L,\alpha}(t|\bm{x}_0,t_0)$ in the $\alpha$ direction for turbulent flows driven by $\bm{f}^\mathrm{(V)}$ with $\mathrm{Re}_\lambda=120$~(black) and $247$~(red), $\bm{f}^\mathrm{(I)}$ with $\mathrm{Re}_\lambda=220$~(blue) and $310$~(orange), and $\bm{f}^\mathrm{(R)}$ with $\mathrm{Re}_\lambda=205$~(green) and $323$~(magenta). The solid lines indicate $\alpha=x$ and the dash-dotted lines indicate $\alpha=z$. The black dashed line indicates $E_{L,\alpha}(\omega)\propto \omega^{-2}$. The dotted lines indicate $\tau_\eta/T_{L,x}$. (b) Compensated spectra $\omega^2E_{L,\alpha}(\omega)/\overline{\epsilon}$ as a function of $\omega\tau_\eta$. The black dashed line shows $\omega^2E_{L,\alpha}(\omega)/\overline{{\epsilon}}=2.1$.}
  
        \label{fig:power_spe}
  \end{figure*}

In this subsection, we demonstrate the universality and non-universality of the Lagrangian velocity spectra through a systematic comparison of turbulent flows with different Reynolds numbers $\mathrm{Re}_\lambda$ and forcing methods $\bm{f}$.
In this study, we define the power spectral density $E_{L,\alpha}(\omega)$ of the Lagrangian velocity $v_{L,\alpha}(t|\bm{x}_0,t_0)$ in the $\alpha$ direction as
\begin{equation}
    E_{L,\alpha}(\omega) = \frac{2\langle|\hat{v}_{L,\alpha}(\omega)|^2\rangle}{T}, \label{eq:Lagrangian_power_spectra}
\end{equation}
where $\hat{v}_{L,\alpha}(\omega)$ is the Fourier coefficient of $v_{L,\alpha}(t|\bm{x}_0,t_0)$ and $T$ is the length of time series of $v_{L,\alpha}(t|\bm{x}_0,t_0)$.
Here, $\langle \cdot \rangle$ denotes the ensemble average over $32^3$ trajectories.
Note that $E_{L,\alpha}(\omega)$ defined as Eq.~\eqref{eq:Lagrangian_power_spectra} satisfies 
\begin{equation}
    \int_0^\infty E_{L,\alpha}(\omega) d\omega = \langle v_{L,\alpha}^2\rangle.%\lim_{T\to \infty}\frac{1}{T}\int_0^T v_{L,\alpha}^2dt.
\end{equation}
Figure~\ref{fig:power_spe}(a) shows $E_{L,x}(\omega)$ for various $\mathrm{Re}_\lambda$ and $\bm{f}$.
Following the Kolmogorov similarity hypothesis, $E_{L,x}(\omega)$ and $\omega$ are normalized by the kinematic viscosity $\nu$ and the Kolmogorov time $\tau_\eta=(\nu/\overline{\epsilon})^{1/2}$, respectively.
We define the Lagrangian timescale $T_{L,\alpha}$ of $v_{L,\alpha}(t|\bm{x}_0,t_0)$ as $T_{L,\alpha}=\int_0^\infty C_{L,\alpha}(\tau)d\tau$, where the autocorrelation function $C_{L,\alpha}(\tau)$ is defined as
\begin{equation}
    C_{L,\alpha}(\tau) = \frac{\langle v_{L,\alpha}(t+\tau|\bm{x}_0,t_0)v_{L,\alpha}(t|\bm{x}_0,t_0)\rangle}{\langle v_{L,\alpha}^2(t|\bm{x}_0,t_0)\rangle }.
\end{equation}
Note that $\langle v_{L,\alpha}\rangle=0$ in the systems considered.
For comparison, we also show $\tau_\eta/T_{L,x}$ in Fig.~\ref{fig:power_spe}(a).
Similarly to the energy spectra $E(k)$~(Fig.~\ref{fig:ene_spe}), $E_{L,x}(\omega)$ also shows the universality for $\omega\gg 1/T_{L,x}$.
The Kolmogorov similarity hypothesis predicts that in the Lagrangian inertial range, $E_{L,x}(\omega)$ obeys the scaling law $E_{L,x}(\omega)=B_0\overline{\epsilon} \omega^{-2}$, where $B_0$ is a universal constant.
To confirm the validity of this Kolmogorov scaling, we show the compensated power spectral density $\omega^2E_{L,x}(\omega)/\overline{\epsilon}$ in Fig.~\ref{fig:power_spe}(b).
We observe that $\omega^2E_{L,x}(\omega)/\overline{\epsilon}$ exhibits a plateau regime with $B_0\simeq 2.1$, thus supporting the Kolmogorov scaling as reported in previous studies~\cite{Mordant2001-td,Yeung2001-rq,Chevillard2005-qe,Yeung2006-nl,Angriman2020-gq}.
However, $E_{L,x}(\omega)$ for $\bm{f}^\mathrm{(V)}$ shows the Kolmogorov scaling in the narrower range than $E_{L,x}(\omega)$ for $\bm{f}^\mathrm{(I)}$ and $\bm{f}^\mathrm{(R)}$ while $E(k)$ almost collapses on the single function at $k\gtrsim k_f$ irrespective of the forcing method~(Fig.~\ref{fig:ene_spe}).
In other words, the extent of the Lagrangian inertial range is more sensitive to the forcing method than that of the Eulerian inertial range.
On the basis of simple dimensional estimations $T_L/\tau_\eta\sim \mathrm{Re}_\lambda$ and $L/\eta\sim \mathrm{Re}_\lambda^{3/2}$ with $L$ being the integral length, this different scaling behavior of the Eulerian and Lagrangian spectra is partly attributed to the fact that the inertial range in $E(k)$ extends faster than $E_L(\omega)$~\cite{Benzi2010-ll}. %(Frisch p.107)
The Kolmogorov scaling is also barely observed in the second-order structure function $S_2(\tau)=\langle [v_{L,\alpha}(t+\tau)-v_{L,\alpha}(t)]^2\rangle$ of the Lagrangian velocity~\cite{Yeung2006-nl,Sawford2011-jp,Falkovich2012-nj,Huang2013-nv,Lanotte2013-ax,Angriman2020-gq}.
Regarding $S_2(\tau)$, large-scale flows are considered to be responsible for the narrow scaling regime in $S_2(\tau)$~\cite{Huang2013-nv,Angriman2022-hs}.
Relatedly, Fig.~\ref{fig:power_spe} has demonstrated that the Lagrangian inertial range of $E_{L,x}(\omega)$ is also greatly influenced by the forcing method~(i.e., a mean flow or the largest-scale flow).
Thus, through a systematic comparison between $E_{L,\alpha}(\omega)$ and $E(k)$ for different forcing methods, we have demonstrated the effect of the largest-scale flow on the Lagrangian velocity.
In Sec.~\ref{subsec:scale-decomposition}, we will  directly demonstrate that the largest-scale flow driven by the external force can contribute to the shrinkage of the Lagrangian inertial range using our scale-decomposition analysis method.

Next, we consider the effect of anisotropy of turbulent flows on the Lagrangian velocity.
Since $\bm{f}^{(\mathrm{V})}$ is an anisotropic forcing function, $E_{L,\alpha}(\omega)$ depends on the direction $\alpha$ of the Lagrangian velocity~\cite{Angriman2020-gq}.
Specifically, $\bm{f}^{(\mathrm{V})}$ does not generate the mean flow in the $z$ direction.
To see the effect of the anisotropy, we show $E_{L,z}(\omega)$ for $\bm{f}^{(\mathrm{V})}$ in Fig.~\ref{fig:power_spe}.
We confirm that $E_{L,z}(\omega)$ also collapses on the universal function for $\omega\gg 1/T_{L,\alpha}$. 
Since the mean flow exists in the $x$ direction, $E_{L,x}(\omega)>E_{L,z}(\omega)$ holds for $\omega\to 0$.
Incidentally, $E_{L,z}(\omega)$ does not exhibit a significant non-universality compared with $E_{L,x}(\omega)$ for $\bm{f}^{(\mathrm{V})}$.
Considering the absence of the mean flow in the $z$ direction, it is reasonable that $E_{L,z}(\omega)$ for $\bm{f}^{(\mathrm{V})}$ behaves in a similar way as $E_{L,x}(\omega)$ for $\bm{f}^{(\mathrm{I})}$ and $\bm{f}^{(\mathrm{R})}$, which do not induce mean flows.
In other words, the mean flow is closely related to the non-universality of $E_{L,\alpha}(\omega)$.

To summarize, for $\omega\gg 1/T_{L,\alpha}$, $E_{L,\alpha}(\omega)$ follows the universal function regardless of the external force, Reynolds number, and velocity direction, whereas, in contrast to $E(k)$, non-universal behavior significantly emerges in the low-frequency regime, depending on the forcing method.
In the next subsection, we apply the scale-decomposition analysis to the Lagrangian velocity to explain the formation mechanism of $E_{L,\alpha}(\omega)$ based on the hierarchy of coherent vortices in turbulence.

% %   --------------------
%   --------------------

\subsection{Scale-decomposition analysis of the Lagrangian velocity\label{subsec:scale-decomposition}}

\begin{figure}
    \centering
    \begin{overpic}[width=0.9\linewidth]{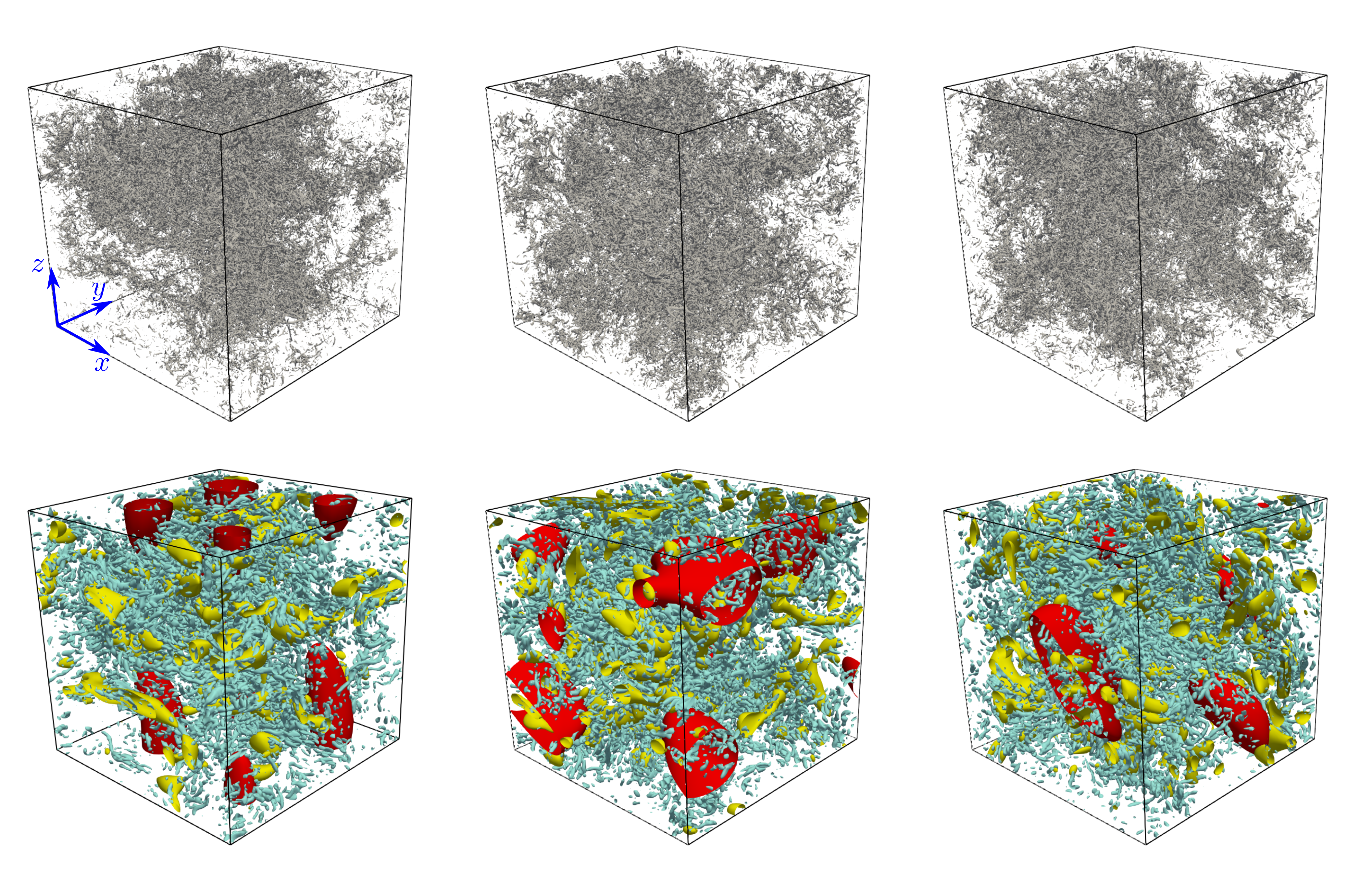} 
          \put(0,60){(a\,i)}
          \put(33,60){(a\,ii)}
          \put(66,60){(a\,iii)}
          \put(0,29){(b\,i)}
          \put(33,29){(b\,ii)}
          \put(66,29){(b\,iii)}

    \end{overpic}
    \caption{Isosurfaces of (a) raw $|\bm{\omega}|^2$ and (b) bandpass-filtered enstrophy $|\bm{\omega}^{(k_c)}|^2$ for (i) $\mathrm{Re}_\lambda=247$ with forcing $\bm{f}^{(\mathrm{V})}$, (ii) $\mathrm{Re}_\lambda=312$ with forcing $\bm{f}^{(\mathrm{I})}$, and (iii) $\mathrm{Re}_\lambda=323$ with forcing $\bm{f}^{(\mathrm{R})}$. In (b), $k_c=\sqrt{2}$~(red), $4\sqrt{2}$~(yellow), and $16\sqrt{2}$~(cyan). In (a) and (b), the threshold value $\mathcal{E}$ of the isosurface
    is set to $\mu +3\sigma$ and $\mu+2\sigma$, respectively. Here, $\mu$ and $\sigma$ denote the spatial average and standard deviation of $|\bm{\omega}|^2$ and $|\bm{\omega}^{(k_c)}|^2$.}
    \label{fig:snapshot}
\end{figure}

\begin{figure}
    \centering
    \begin{overpic}[width=0.9\linewidth]{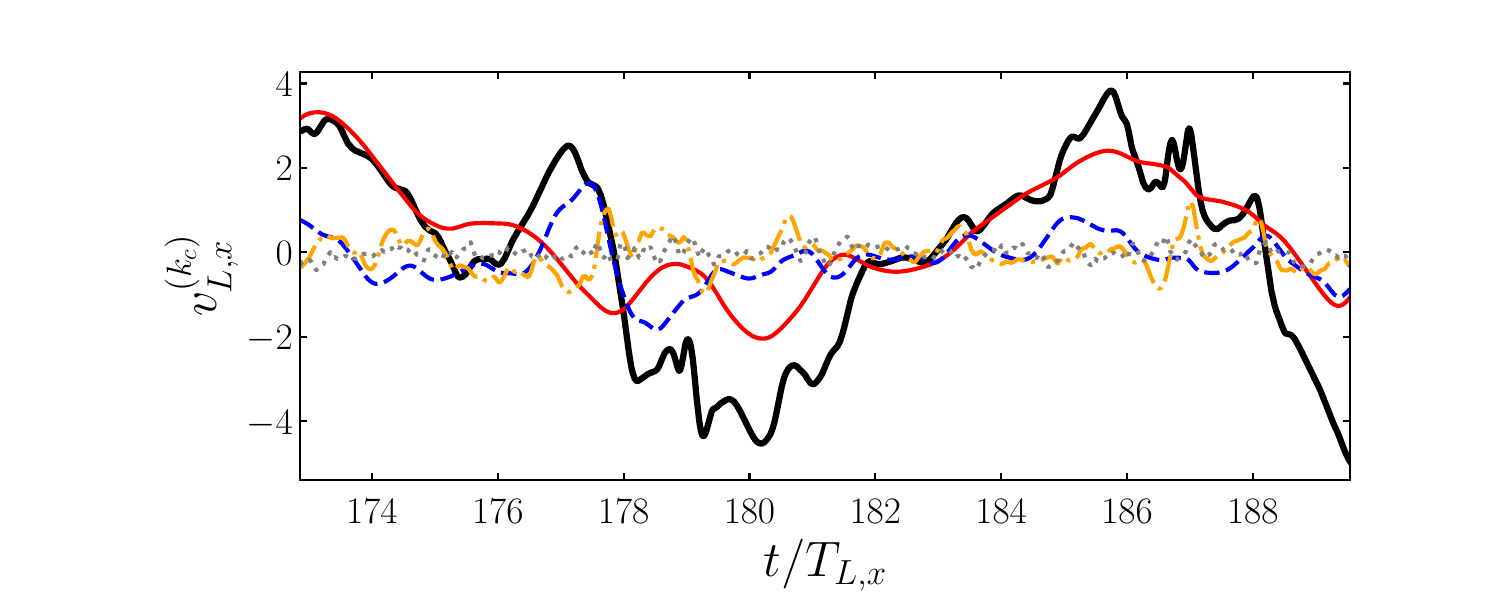} 
    \end{overpic}
    \caption{Bandpass-filtered Lagrangian velocity $v_{L,x}^{(k_c)}(t|\bm{x}_0,t_0)$ in the $x$ direction as a function of time $t/T_{L,x}$ at $\mathrm{Re}_\lambda =247$ with forcing $\bm{f}^{(\mathrm{V})}$ for $k_c=k_f(=\sqrt{2})$~(red line), $4k_f$~(blue dashed line), $16k_f$~(orange dash-dotted line), and $64k_f$~(gray dotted line). The black solid line shows the raw Lagrangian velocity $v_{L,x}(t|\bm{x}_0,t_0)$ in the $x$ direction.}
    \label{fig:lag_bp_vel_time}
\end{figure}

In Sec.~\ref{subsec:power_spe}, a comparison of $E_{L,\alpha}(\omega)$ for different $\mathrm{Re}_\lambda$ and $\bm{f}$ has demonstrated that $E_{L,\alpha}(\omega)$ shows a significant non-universality compared with $E(k)$ due to the difference in the forcing method, although $E_{L,\alpha}(\omega)$ collapses on a universal function in the high-frequency regime.
In this subsection, we develop a scale-decomposition analysis method for the Lagrangian velocity to understand the formation mechanism of $E_{L,\alpha}(\omega)$ in terms of the hierarchy of coherent vortices in turbulence.
Although several studies focus on the time series analysis of the Lagrangian velocity~\cite{Huang2013-nv,Angriman2022-hs}, we apply the scale-decomposition to the Eulerian velocity field $\bm{u}(\bm{x},t)$ to relate the Eulerian spatial characteristics and the Lagrangian dynamical characteristics~\cite{Koide2024-lc}.
Specifically, we first define the velocity $\bm{u}^{(k_c)}(\bm{x},t)$ at wave number $k_c$ as the velocity obtained using the Fourier bandpass filter with passband $[k_c/\sqrt{2},\sqrt{2}k_c]$~\cite{Goto2017-im,Hirota2020-jm}.
Figure~\ref{fig:snapshot} shows the isosurfaces of the enstrophy $|\bm{\omega}|^2$ and the bandpass-filtered enstrophy $|\bm{\omega}^{(k_c)}|^2$ for different forcing methods.
With the aid of the bandpass filter, we can extract vortical structures at each scale.
Using $\bm{u}^{(k_c)}(\bm{x},t)$, we define the Lagrangian velocity $\bm{v}_L^{(k_c)}(t|\bm{x}_0,t_0)$ at wave number $k_c$ as
\begin{equation}
    \bm{v}_L^{(k_c)}(t|\bm{x}_0,t_0) = \bm{u}^{(k_c)}(\bm{x}_L(t|\bm{x}_0,t_0),t).\label{eq:decomposed_Lagrangian}
\end{equation}
In Eq.~\eqref{eq:decomposed_Lagrangian}, we compute $\bm{u}^{(k_c)}(\bm{x}_L(t|\bm{x}_0,t_0),t)$ by appling the trilinear interpolation to $\bm{u}^{(k_c)}(\bm{x},t)$.
Since $\bm{u}(\bm{x}_L(t|\bm{x}_0,t_0),t)=\sum_{k_c}\bm{u}^{(k_c)}(\bm{x}_L(t|\bm{x}_0,t_0),t)$,
Eq.~\eqref{eq:decomposed_Lagrangian} allows us to decompose $\bm{v}_L(t|\bm{x}_0,t_0)(=\bm{u}(\bm{x}_L(t|\bm{x}_0,t_0),t))$ into the contribution from each scale:
\begin{equation}
    \bm{v}_L(t|\bm{x}_0,t_0) = \sum_{k_c} \bm{v}_L^{(k_c)}(t|\bm{x}_0,t_0). \label{eq:Lagrangian_velocity_decomposition}
\end{equation}
Note that the trajectory $\bm{x}_L(t|\bm{x}_0,t_0)$ of fluid particles remains unaffected by the scale decomposition; the time evolution of $\bm{x}_L(t|\bm{x}_0,t_0)$ is based on the raw velocity field $\bm{u}(\bm{x},t)$.
Figure~\ref{fig:lag_bp_vel_time} shows the time series of $v_{L,x}^{(k_c)}(t|\bm{x}_0,t_0)$ for various $k_c$ at $\mathrm{Re}_\lambda =247$ with forcing $\bm{f}^{(\mathrm{V})}$.
We observe that $v_{L,x}^{(k_c)}(t|\bm{x}_0,t_0)$ at smaller $k_c$ tends to exhibit larger and slower fluctuations, which is consistent with the conventional picture of turbulence.
In the following, we directly evaluate the contribution of vortices with different sizes to $E_{L,\alpha}(\omega)$ using scale decomposition analysis~[Eq.~\eqref{eq:Lagrangian_velocity_decomposition}].

Before presenting our results, it is worth noting that Machicoane and Volk~\cite{Machicoane2016-fc} decomposed the velocity of inertial particles into the local mean flow value at the particle position and the fluctuating velocity.
This method enables us to disentangle the contributions of mean flows and turbulence to the Lagrangian velocity.
However, since we are aimed at relating the Lagrangian velocity statistics to the hierarchy of coherent vortices with different length scales, the present study adopts our scale-decomposition method~[Eq.~\eqref{eq:Lagrangian_velocity_decomposition}].
In addition, Eq.~\eqref{eq:Lagrangian_velocity_decomposition} is based on the instantaneous velocity field, in contrast to the subtraction of the mean flow contribution, which requires the average of the Eulerian velocity field over different snapshots.
Thus, our scale-decomposition method allows us to take into account the temporal fluctuations of the largest-scale flows, which have been reported for $\bm{f}^{(\mathrm{V})}$~\cite{Yasuda2014-tm,Goto2017-im}.

%   --------------------
%   --------------------
%   --------------------
%   --------------------

\begin{figure}
    \centering
    \begin{overpic}[width=0.5\linewidth]{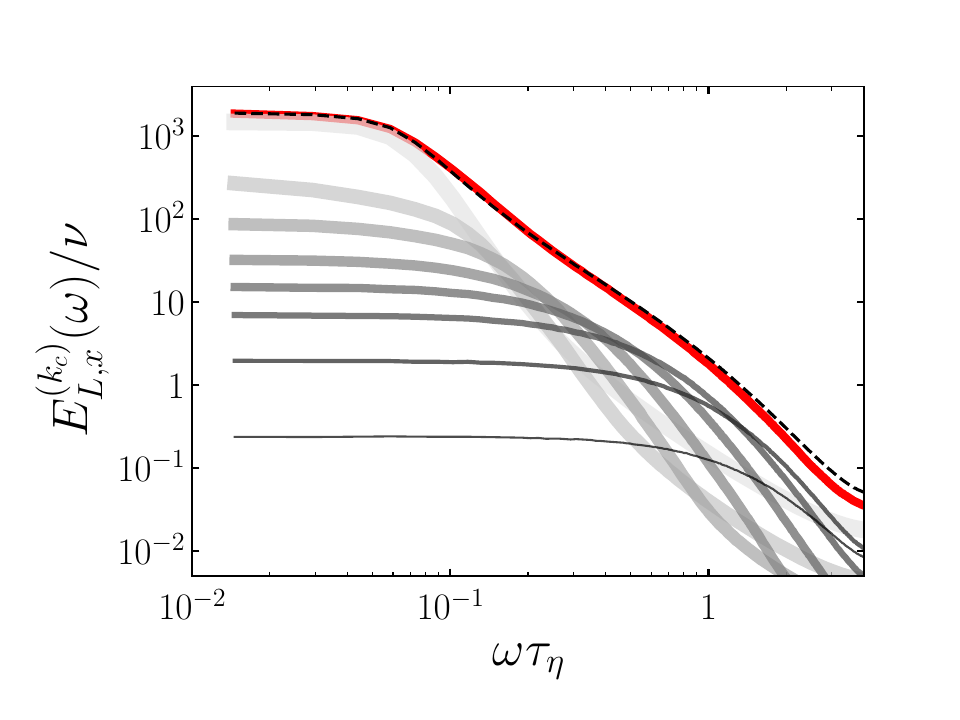} 
    \end{overpic}
    \caption{Power spectral density $E_{L,x}^{(k_c)}(\omega)$ of the bandpass-filtered Lagrangian velocity $v_{L,x}^{(k_c)}(t|\bm{x}_0,t_0)$ in the $x$ direction for $\mathrm{Re}_\lambda=247$ with forcing $\bm{f}^{(\mathrm{V})}$. From thicker~(and lighter) to thinner~(and darker) lines, $k_c=k_f(=\sqrt{2}),~2k_f,~4k_f,~8k_f,~16k_f,~32k_f,~64k_f$, and $128k_f$. The black dashed line and the red line show $\sum_{k_c}E_{L,x}^{(k_c)}(\omega)$ and $E_{L,x}(\omega)$, respectively.}
    \label{fig:power_spe_bp_validation}
\end{figure}

We directly relate the hierarchical structures in turbulence to the power spectral density of the Lagrangian velocity on the basis of the scale decomposition in Eq.~\eqref{eq:Lagrangian_velocity_decomposition}.
With Eqs.~\eqref{eq:Lagrangian_power_spectra} and \eqref{eq:Lagrangian_velocity_decomposition}, $E_{L,\alpha}(\omega)$ is expressed as 
\begin{equation}
    E_{L,\alpha}(\omega) = \sum_{k_c}\frac{2\langle|\hat{v}_{L,\alpha}^{(k_c)}(\omega)|^2\rangle}{T} + \sum_{k_c,k_c^\prime(\neq k_c)}\frac{4\langle\Re\{\hat{v}_{L,\alpha}^{(k_c)}(\omega){\hat{v}_{L,\alpha}^{(k_c^\prime)*}} (\omega)\}\rangle}{T}.\label{eq:Lagrangian_power_spectra_decomposition}
\end{equation}
Here, $(\cdot)^*$ denotes the complex conjugate, and $\operatorname{Re}\{\cdot\}$ denotes the real part of a complex number.
We show $E_{L,x}(\omega)$ and the first term on the right-hand side of Eq.~\eqref{eq:Lagrangian_power_spectra_decomposition} for $\mathrm{Re}_\lambda=247$ with forcing $\bm{f}^{(\mathrm{V})}$ in Fig.~\ref{fig:power_spe_bp_validation} with the red line and the black dashed line, respectively. 
Since this comparison shows that the second term on the right-hand side of Eq.~\eqref{eq:Lagrangian_power_spectra_decomposition} has a negligible effect on $E_{L,x}(\omega)$, it is sufficient to consider the contribution from each scale separately.
Thus, we define the contribution $E_{L,\alpha}^{(k_c)}(\omega)$ of flows at wave number $k_c$ to $E_{L,\alpha}(\omega)$ as
\begin{equation}
    E_{L,\alpha}^{(k_c)}(\omega) = \frac{2\langle|\hat{v}_{L,\alpha}^{(k_c)}(\omega)|^2\rangle}{T}.
\end{equation}
We show $E_{L,x}^{(k_c)}(\omega)$ for different $k_c$ in Fig.~\ref{fig:power_spe_bp_validation} with the gray lines, indicating that $E_{L,x}(\omega)$ is formed by the contribution from vortices with different sizes.
In the following, we will reveal the role of vortices with different sizes in the formation of $E_{L,\alpha}(\omega)$ by comparing $E_{L,\alpha}^{(k_c)}(\omega)$ for different turbulent flows.
%   --------------------
%   --------------------

\begin{figure*}
    \centering
    \begin{overpic}[width=0.5\linewidth]{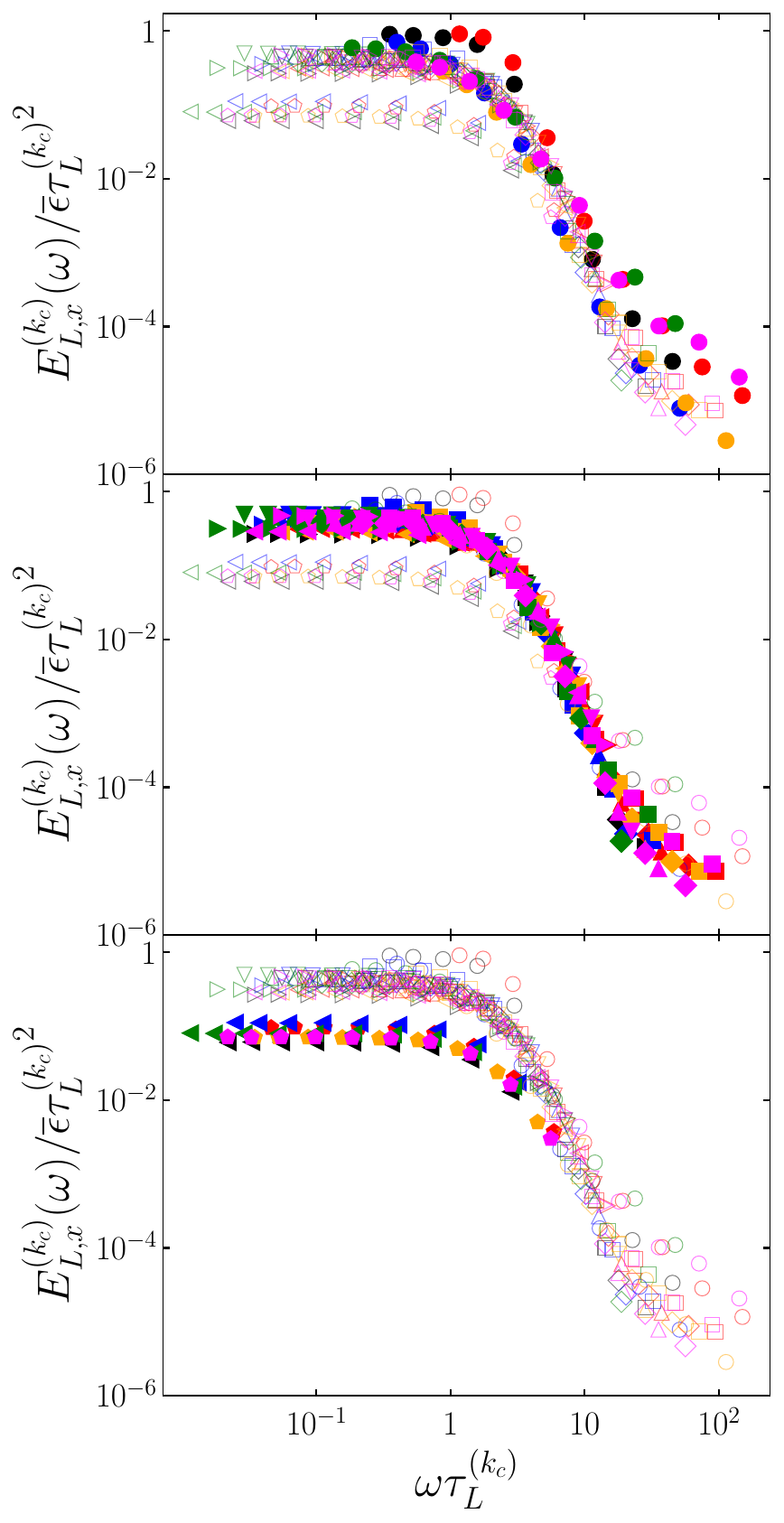} 
          \put(-0,98){(a)}
          \put(-0,67){(b)}
          \put(-0,36){(c)}

    \end{overpic}

    \caption{Power spectral density $E_{L,x}^{(k_c)}(\omega)$ of the bandpass-filtered Lagrangian velocity $v_{L,x}^{(k_c)}(t|\bm{x}_0,t_0)$ in the $x$ direction normalized by $\overline{\epsilon}$ and the Lagrangian timescale $\tau_L^{(k_c)}$ of the flows at wave number $k_c$ for turbulent flows driven by $\bm{f}^\mathrm{(V)}$ with $\mathrm{Re}_\lambda=120$~(black) and $247$~(red), $\bm{f}^\mathrm{(I)}$ with $\mathrm{Re}_\lambda=220$~(blue) and $310$~(orange), and $\bm{f}^\mathrm{(R)}$ with $\mathrm{Re}_\lambda=205$~(green) and $323$~(magenta). Different symbols correspond to different $k_c$: $\circ$, $k_c=\sqrt{2}$; $\square$, $2\sqrt{2}$; $\diamond$, $4\sqrt{2}$; $\triangle$, $8\sqrt{2}$; $\triangledown$, $16\sqrt{2}$; $\triangleright$, $32\sqrt{2}$; $\triangleleft$, $64\sqrt{2}$; $\pentago$, $128\sqrt{2}$. Results in (a) the energy-containing range~($k_c=\sqrt{2}$), (b) inertial range~($\sqrt{2}<k_c\lesssim 0.3\eta^{-1}$), and (c) dissipation range~($k_c\gtrsim 0.3\eta^{-1}$) are emphasized by filled symbols.}
  
        \label{fig:power_spe_bp}
  \end{figure*}

\begin{figure}
    \centering
    \begin{overpic}[width=0.5\linewidth]{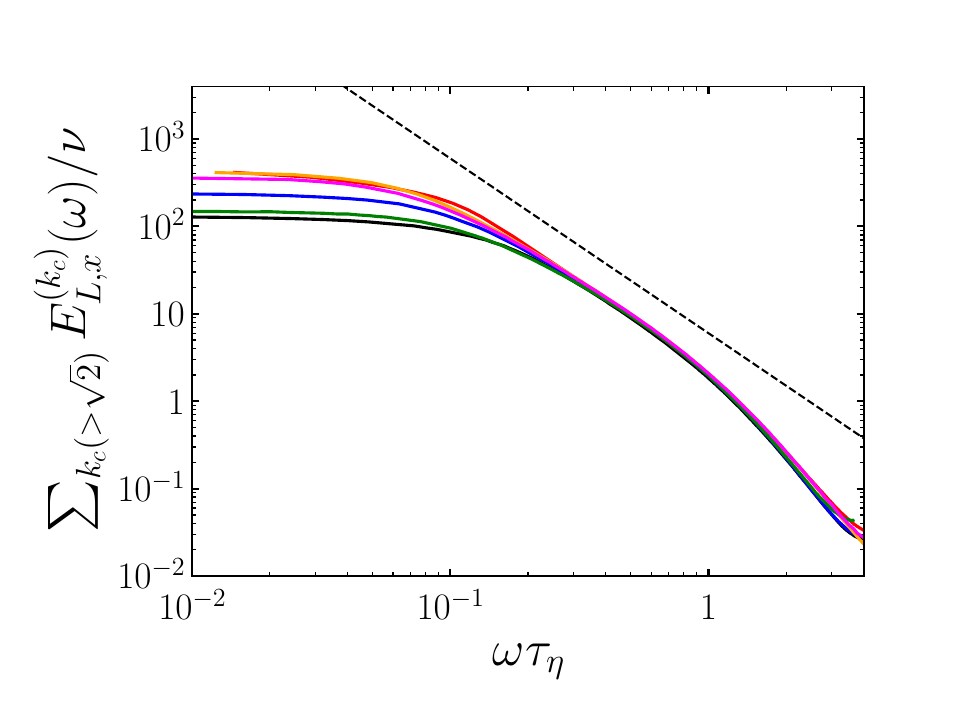} 
    \end{overpic}
    \caption{Superposition $\sum_{k_c(>\sqrt{2})} E_{L,x}^{(k_c)}(\omega)$ of the power spectral density $E_{L,x}^{(k_c)}(\omega)$ of the bandpass-filtered Lagrangian velocity $v_{L,x}^{(k_c)}(t|\bm{x}_0,t_0)$ in the $x$ direction, where the contribution from the forcing scale $k_c=\sqrt{2}(\simeq k_f)$ is subtracted. The colors are the same as in Fig.~\ref{fig:power_spe}. The dashed line indicates a power law with the exponent $-2$.}
    \label{fig:power_spe_bp_sum}
\end{figure}

To explore the origin of the universality and non-universality of $E_{L,\alpha}(\omega)$, we compare $E_{L,x}^{(k_c)}(\omega)$ for different turbulent flows.
Figure~\ref{fig:power_spe_bp} shows $E_{L,x}^{(k_c)}(\omega)$ for various $k_c$.
Note that following the Kolmogorov similarity hypothesis, $\omega$ and $E_{L,x}^{(k_c)}(\omega)$ are normalized by $\overline{\epsilon}$ and the Lagrangian timescale $\tau_L^{(k_c)}=\overline{\epsilon}^{-1/3}k_c^{-2/3}$ of vortices at wave number $k_c$.
Although each panel in Fig.~\ref{fig:power_spe_bp} shows the same data, the results in the energy-containing range~($k_c=\sqrt{2}$), inertial range~($\sqrt{2}<k_c\lesssim 0.3\eta^{-1}$), and dissipation range~($k_c\gtrsim 0.3\eta^{-1}$) are highlighted by filled symbols in (a), (b), and (c), respectively.
We find that for the inertial range $k_f \lesssim k_c \lesssim 0.3 \eta^{-1}$, $E_{L,x}^{(k_c)}(\omega)$~[Fig.~\ref{fig:power_spe_bp}(b)] collapses on a single function regardress of $\bm{f}$ and $\mathrm{Re}_\lambda$.
The collapse indicates that small-scale vortices in the inertial range induce the self-similar dynamics of the Lagrangian velocity.
These universal dynamics of the small-scale vortices yield the Lagrangian inertial range in $E_{L,x}(\omega)$~(Fig.~\ref{fig:power_spe}).
In addition, it is worth emphasizing that an appropriate timescale for the collapse is $\tau_L^{(k_c)}$, rather than the Eulerian timescale $\tau_E^{(k_c)}$ defined as $\tau_E^{(k_c)}=1/(k_cU_0)\propto k_c^{-1}$ with $U_0$ being the sweeping velocity.
The relevance of $\tau_L^{(k_c)}$ in the Lagrangian velocity is consistent with the previous results of the correlation function of the Fourier coefficient of the Lagrangian velocity~\cite{Gotoh1993-qm,Kaneda1999-rl,Matsumoto2021-dg}.
In contrast, for $k_c\simeq k_f$, $E_{L,x}^{(k_c)}(\omega)$~[Fig.~\ref{fig:power_spe_bp}(a)] deviates from the universal function observed for $E_{L,x}^{(k_c)}(\omega)$ at $k_f\lesssim k_c\lesssim 0.3\eta^{-1}$ and strongly depends on $\bm{f}$.
Especially for $\bm{f}^{(\mathrm{V})}$, $E_{L,x}^{(k_c)}(\omega)$ at $k_c= k_f$ has a relatively large value at $\omega\tau_L^{(k_c)}\simeq 1$, which results in the significant non-universality of $E_{L,\alpha}(\omega)$~(Fig.~\ref{fig:power_spe}). 
We, therefore, conclude that the Lagrangian inertial range can be narrower compared with $E(k)$ because the dynamics of the largest-scale flows is non-universal depending on the external force and can contaminate the universal power spectra formed by small-scale vortices, whose characteristic velocity amplitudes and timescales obey the Kolmogorov similarity hypothesis.
In fact, as shown in Fig.~\ref{fig:power_spe_bp_sum}, a broader Lagrangian inertial range is observed for $\sum_{k_c(>\sqrt{2})} E_{L,x}^{(k_c)}(\omega)$, where the contribution from the forcing scale $k_c=\sqrt{2}(\simeq k_f)$ is subtracted.
Incidentally, while $E_{L,x}^{(k_c)}(\omega)$~[Fig.~\ref{fig:power_spe_bp}(c)] at $k_c\gtrsim 0.3\eta^{-1}$~(i.e., wave numbers in the dissipation range) falls below the master curve for the inertial range, $\bm{f}$ and $\mathrm{Re}_\lambda$ appear to have little effect on $E_{L,x}^{(k_c)}(\omega)$ at $k_c\gtrsim 0.3\eta^{-1}$.
These tendencies are consistent with the fact that for $\omega\tau_\eta\gtrsim 1$, $E_{L,x}(\omega)$ rapidly decays and almost collapses on a single function~(Fig.~\ref{fig:power_spe}).
However, since the accurate evaluation of $E_{L,x}(\omega)$ in the high-frequency range requires more careful analysis~\cite{Yeung1989-ak}, further studies need to be performed in order to discuss the universality of the Lagrangian dynamics originating from flows in the dissipation range.

% %   --------------------
%   --------------------
%   --------------------
%   --------------------

\begin{figure}
    \centering
    \begin{overpic}[width=0.5\linewidth]{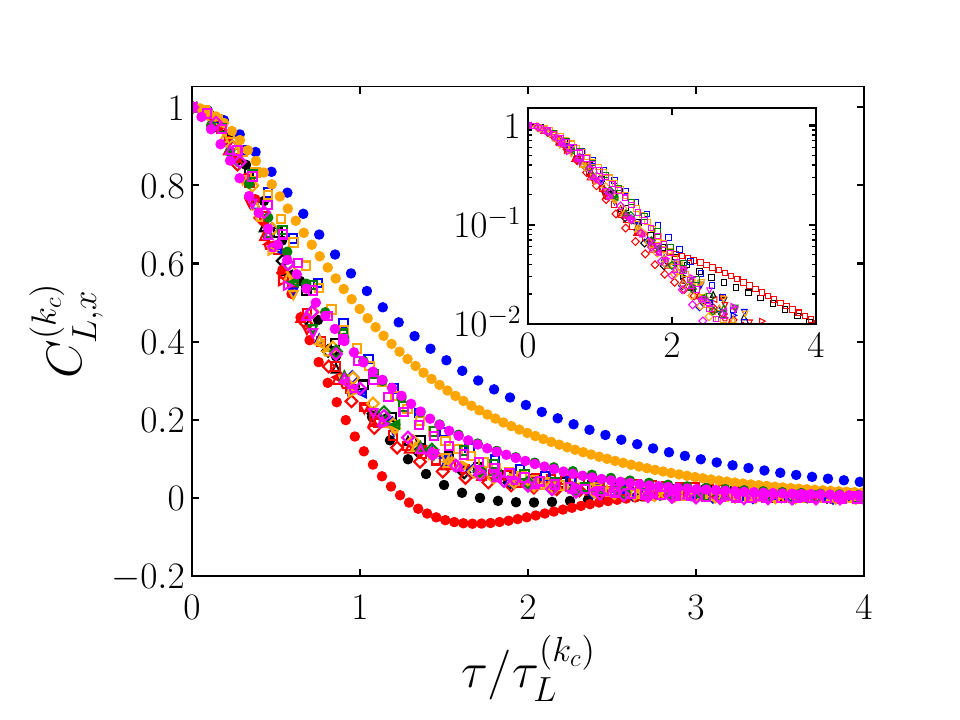} 
    \end{overpic}
    \caption{Autocorrelation function $C_{L,x}^{(k_c)}(\tau)$ of the bandpass-filtered Lagrangian velocity $v_{L,x}^{(k_c)}(t|\bm{x}_0,t_0)$ in the $x$ direction as a function of $\tau/\tau_L^{(k_c)}$. The colors and symbols are the same as in Fig.~\ref{fig:power_spe_bp}(a) with $C_{L,x}^{(k_c)}(\tau)$ at $k_c=\sqrt{2}$ shown by filled symbols. The inset shows the semi-logarithmic plot of $C_{L,x}^{(k_c)}(\tau)$ except for $k_c=\sqrt{2}$.}
    \label{fig:bp_corr}
\end{figure}

Before closing this subsection, we refer to the autocorrelation function $C_{L,\alpha}(\tau)$ of the Lagrangian velocity, which corresponds to the Fourier transform of $E_{L,\alpha}(\omega)$.
Similarly to $E_{L,\alpha}^{(k_c)}(\omega)$, we introduce the scale-decomposed autocorrelation function $C_{L,\alpha}^{(k_c)}(\tau)$ defined as 
\begin{equation}
    C_{L,\alpha}^{(k_c)}(\tau) = \frac{\langle v_{L,\alpha}^{(k_c)}(t+\tau|\bm{x}_0,t_0)v_{L,\alpha}^{(k_c)}(t|\bm{x}_0,t_0)\rangle}{\langle {v_{L,\alpha}^{(k_c)}}^2(t|\bm{x}_0,t_0)\rangle }.
\end{equation}
Figure~\ref{fig:bp_corr} shows $C_{L,x}^{(k_c)}(\tau)$ as a function of $\tau/\tau_L^{(k_c)}$ for different turbulent flows.
As observed for $E_{L,x}^{(k_c)}(\omega)$, $C_{L,x}^{(k_c)}(\tau)$ exhibits the various behaviors depending on $\bm{f}$ for $k_c=\sqrt{2}(\simeq k_f)$, whereas $C_{L,x}^{(k_c)}(\tau)$ collapses on a single function for $k_f\lesssim k_c\lesssim 0.3\eta^{-1}$, which is more evident in the semi-logarithmic plot of $C_{L,x}^{(k_c)}(\tau)$ in the inset of Fig.~\ref{fig:bp_corr}.
It may be worth mentioning that for $\bm{f}^\mathrm{(V)}$, $C_{L,x}^{(k_c)}(\tau)$ at $k_c=2\sqrt{2}$ deviates from the master curve at $\tau/\tau_L^{(k_c)}\gtrsim 2$, which is probably due to a remaining effect of the largest-scale flows at $k_c=\sqrt{2}$.
Indeed, this deviation almost disappears for $k_c>2\sqrt{2}$.
It would be interesting to explore the functional form of the obtained master curve, which is left for a future study.
%   --------------------
%   --------------------

\subsection{Comparison between the power spectral density of the Eulerian and Lagrangian velocity}

\begin{figure*}
    \centering
        \begin{tabular}{c}
        \begin{minipage}{0.5\hsize}
            \begin{overpic}[width=1\linewidth]{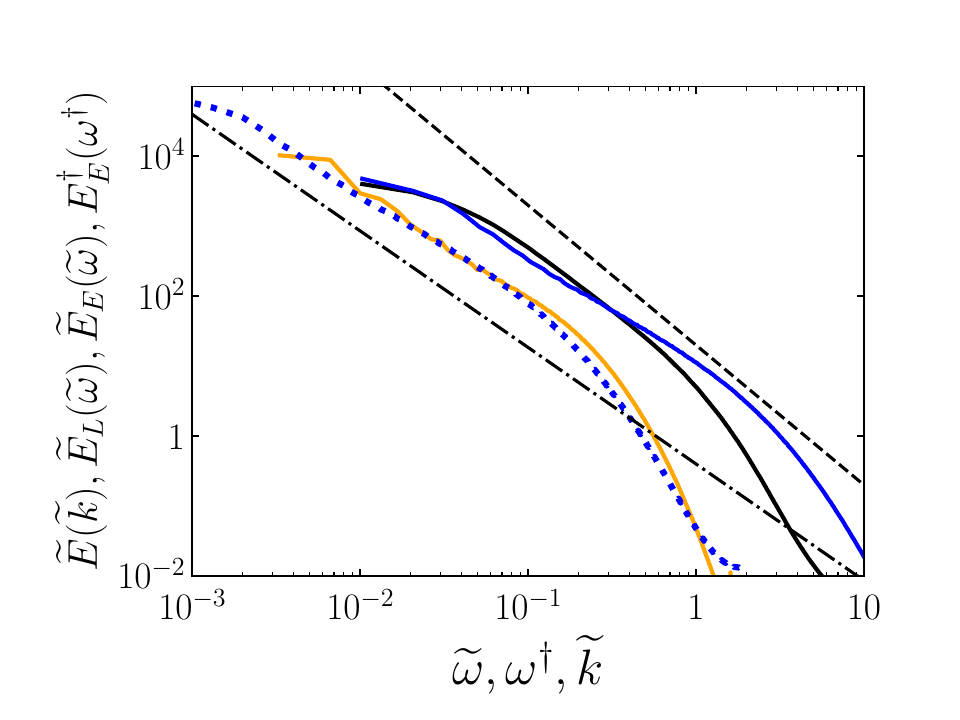}
                \linethickness{3pt}
          \put(0,63){(a)}

            \end{overpic}
        \end{minipage}
        \begin{minipage}{0.5\hsize}
            \begin{overpic}[width=1\linewidth]{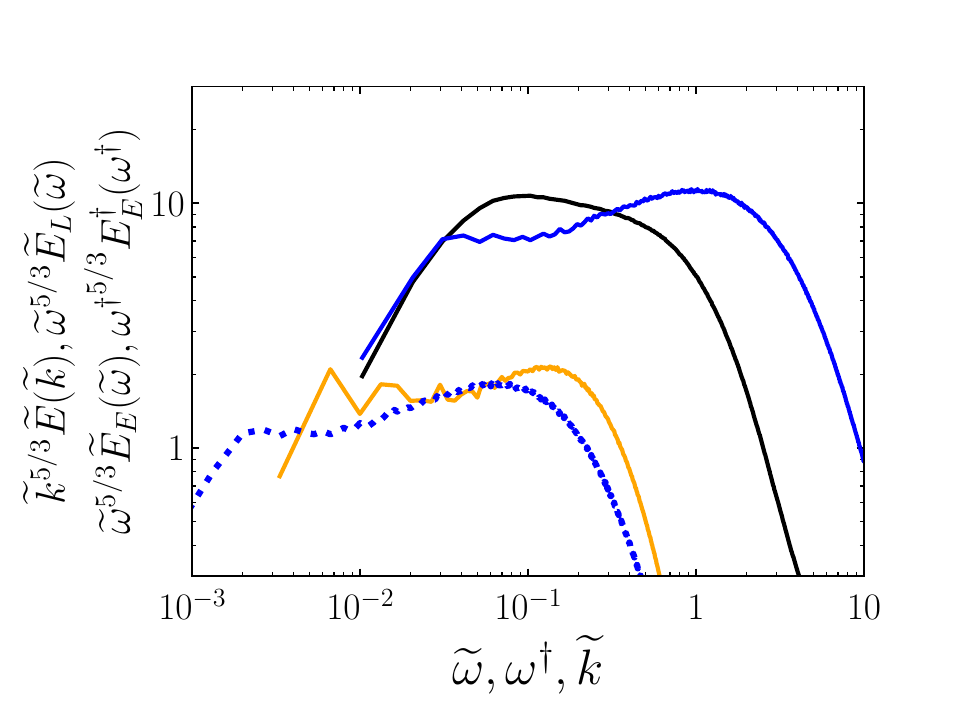}
                \linethickness{3pt}
          \put(0,63){(b)}

            \end{overpic}
        \end{minipage}
        \end{tabular}
    \caption{(a) Comparison of the energy spectrum $E(k)$~(orange), the Lagrangian power spectrum $E_L(\omega)$~(black), and the Eulerian power spectrum $E_E(\omega)$~(blue) for $\mathrm{Re}_\lambda = 310$ with forcing $\bm{f}^\mathrm{(I)}$. Here, solid lines indicate the normalized value $\widetilde{(\cdot)}$ by $\nu$ and $\overline{\epsilon}$. The blue dotted line indicates $E_E^\dag(\omega^\dag)$ as a function of $\omega^\dag$, where $\omega^\dag=\omega\eta/u^\prime$ and $E_E^\dag=u^\prime E_E/(\overline{\epsilon}^{1/4}\nu^{5/4})$. The dashed and dash-dotted lines indicate power laws with the exponents $-2$ and $-5/3$, respectively. (b) Compensated spectra $\widetilde{k}^{5/3}\widetilde{E}(\widetilde{k})$, $\widetilde{\omega}^{5/3}\widetilde{E}_L(\widetilde{\omega})$, $\widetilde{\omega}^{5/3}\widetilde{E}_E(\widetilde{\omega})$, and $\omega^{\dag{5/3}}E_E^\dag(\omega^\dag)$.}
  
        \label{fig:all_spe}
  \end{figure*}

This subsection is aimed at revealing that the conspicuous features of $E_{L,\alpha}(\omega)$ shown in this study are specific to the Lagrangian velocity through a comparison with the power spectral density $E_E(\omega)$ of the Eulerian velocity, which is also obtained from the time series of the velocity.
For a given time series of the Eulerian velocity $u_{E,\alpha}(t|\bm{x})$ in the $\alpha$ direction at a fixed position $\bm{x}$, $E_{E,\alpha}(\omega)$ is defined as
\begin{equation}
    E_{E,\alpha}(\omega) = \frac{2\overline{|\hat{u}_{E,\alpha}(\omega)|^2}}{T}, \label{eq:Euler_psd}
\end{equation}
where $\hat{u}_{E,\alpha}(\omega)$ is the Fourier coefficient of $u_{E,\alpha}(t|\bm{x})$ and $\overline{(\cdot)}$ denotes the spatial average, which is justified by the homogeneity of turbulent flows in the case of $\bm{f}^{\mathrm{(I)}}$ and $\bm{f}^{\mathrm{(R)}}$.
Figure~\ref{fig:all_spe}(a) compares $\widetilde{E}(\widetilde{k})$, $\widetilde{E}_{E}(\widetilde{\omega})$, and $\widetilde{E}_{L}(\widetilde{\omega})$ of turbulent flows driven by $\bm{f}^{(\mathrm{I})}$ with $\mathrm{Re}_\lambda=310$.
Here, $\widetilde{(\cdot)}$ denotes a nondimensionalized quantity by $\overline{\epsilon}$ and $\nu$.
Since $\bm{f}^{(\mathrm{I})}$ generates isotropic turbulence, we consider $E_{L}(\omega)=\sum_\alpha E_{L,\alpha}(\omega)$ and $E_{E}(\omega)=\sum_\alpha E_{E,\alpha}(\omega)$.
To compare $E(k)$ and $E_{E}(\omega)$, we also show $E_E^\dag = u^\prime E_E/(\overline{\epsilon}^{1/4}\nu^{5/4})$ as a function of $\omega^\dag =\omega\eta/u^\prime$ on the assumption that large-scale flows sweep the small-scale structures of turbulence~\cite{Tennekes1975-gz}, where $u^\prime$ denotes the root-mean-square of the velocity fluctuation.
Specifically, we assume the relation $k \sim \omega/u^\prime$, thus leading to $E(k) \sim u^\prime E_E(\omega)$, that is, $\widetilde{E}(\widetilde{k})\sim E_{E}^\dag(\omega^\dag)$.
Indeed, the orange solid and the blue dotted lines in Fig.~\ref{fig:all_spe}(a) demonstrate the validity of $\widetilde{E}(\widetilde{k})\sim E_{E}^\dag(\omega^\dag)$, although there is a slight difference between them probably because of the approximation of the sweeping velocity by $u^\prime$.
To closely examine the power law, we also show the same spectra compensated by a power law with an exponent $-5/3$ in Fig.~\ref{fig:all_spe}(b).
We confirm that $E_{E}(\omega)$ exhibits a scaling law $E_{E}(\omega)\propto \omega^{-5/3}$ in a similar way as $E(k)\propto k^{-5/3}$.
Since $\bm{f}^{(\mathrm{I})}$ does not generate a mean flow, the similarity between $\widetilde{E}(\widetilde{k})$ and $E_{E}^\dag(\omega^\dag)$ indicates that the sweeping of small-scale vortices by large-scale flows is dominant in $E_{E}(\omega)$ even for turbulence without a mean flow~\cite{Tennekes1975-gz}, which has been confirmed by the previous studies~\cite{Sanada1992-yb,Tsinober2001-md}.
Thus, $E_{E}^\dag(\omega^\dag)$ mainly reflects the spatial structure of small-scale vortices swept by the largest-scale flows.
Regarding $E_{L}(\omega)$, we find that $\widetilde{E}_L(\widetilde{\omega})$ is slightly steeper than $\widetilde{E}_E(\widetilde{\omega})$ in the Lagrangian inertial range.
This comparison between $\widetilde{E}_L(\widetilde{\omega})$ and $\widetilde{E}_E(\widetilde{\omega})$ also supports the scaling law $E_{L,\alpha}(\omega)\propto \omega^{-2}$ observed in Fig.~\ref{fig:power_spe}.

% %   --------------------
%   --------------------

\begin{figure}
    \centering
    \begin{overpic}[width=0.5\linewidth]{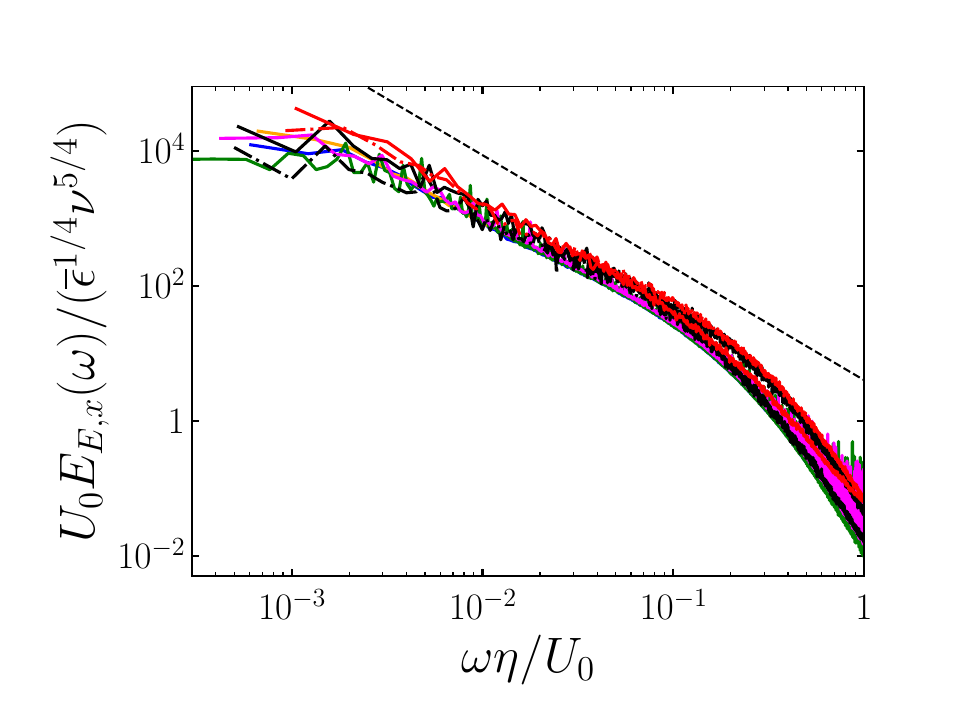} 
    \end{overpic}
    \caption{Power spectral density $E_{E,x}(\omega)$ of the Eulerian velocity $u_{E,x}(t|\bm{x})$ in the $x$ direction at a fixed position $\bm{x}$. The colors are the same as in Fig.~\ref{fig:power_spe}. For $\bm{f}^\mathrm{(V)}$, the solid and dash-dotted lines indicate $E_{E,x}(\omega)$ at $(x,y)=(\pi,\pi)$ and $(x,y)=(\pi,3\pi/2)$, respectively. The dashed line indicates $E_{E,x}(\omega) \propto \omega^{-5/3}$.}

    \label{fig:euler_spe}
\end{figure}

In Sec.~\ref{subsec:power_spe}, we have demonstrated that the power spectral density $E_{L,\alpha}(\omega)$ of the Lagrangian velocity exhibits a significant non-universality originating from the external force, thus leading to the narrow Lagrangian inertial range.
To demonstrate that this characteristic is specific to the Lagrangian velocity, we also examine the universality and non-universality of $E_{E,x}(\omega)$, which is also defined in the frequency domain.
Figure~\ref{fig:euler_spe}(a) shows $E_{E,x}(\omega)$ for turbulent flows with different $\bm{f}$ and $\mathrm{Re}_\lambda$.
We note that in the case of $\bm{f}^{(\mathrm{V})}$, the statistics of $u_{E,\alpha}(t|\bm{x})$ depends on the position $\bm{x}$ due to the nonuniform velocity field caused by $\bm{f}^{(\mathrm{V})}$.
Thus, we evaluate $E_{E,x}(\omega)$ for $\bm{f}^{(\mathrm{V})}$ by taking the average over $z$ for fixed $(x,y)$ in Eq.~\eqref{eq:Euler_psd}.
Specifically, we choose $(x,y)=(\pi,\pi)$, where the amplitude of the mean velocity $\overline{U} \simeq 0$, and $(x,y)=(\pi,3\pi/2)$, where $\overline{U} \simeq 0.8 u^\prime$, to investigate the effect of the mean flow.
Considering the sweeping effect by large-scale flows~\cite{Tennekes1975-gz}, we show the normalized values $\omega\eta/U_0$ and $U_0 E_E(\omega)/(\epsilon^{1/4}\nu^{5/4})$ in Fig.~\ref{fig:euler_spe}, where $U_0$ is the characteristic sweeping velocity.
We use $u^\prime$ as $U_0$ except for $E_{E,x}(\omega)$ at $(x,y)=(\pi,3\pi/2)$ with forcing $\bm{f}^{(\mathrm{V})}$, where we set $U_0 =\overline{U}$.
We find that $E_{E,x}(\omega)$ exhibits a universal behavior in a broad range, which is consistent with the Taylor hypothesis~\cite{Taylor1938-bu,Tennekes1975-gz} and the universality of $E(k)$~(Fig.~\ref{fig:ene_spe}).
In addition, the mean flow has little influence on $E_{E,x}(\omega)$ in the case of $\bm{f}^{(\mathrm{V})}$.
Therefore, we conclude that the significant non-universality of $E_{L,\alpha}(\omega)$ in the low-frequency regime~(Fig.~\ref{fig:power_spe}) is inherent in the Lagrangian velocity.

%   --------------------
%   --------------------

\section{\label{sec:Conclusion}Conclusions}

We have demonstrated the origin of the universality and non-universality of the power spectral density of the Lagrangian velocity in terms of the hierarchy of coherent vortices in turbulence.
For this purpose, we have conducted DNS of turbulent flows in a periodic cube for various Reynolds numbers $\mathrm{Re}_\lambda$ and forcing methods $\bm{f}$.
As is well known, the energy spectrum $E(k)$ follows a single universal curve at $k\gtrsim k_f$ regardress of $\mathrm{Re}_\lambda$ and $\bm{f}$, and the non-universality due to $\bm{f}$ appears only at $k\lesssim k_f$~(Fig.~\ref{fig:ene_spe}).
In contrast, the power spectral density $E_{L,\alpha}(\omega)$ of the Lagrangian velocity has a narrow inertial range, where the Kolmogorov scaling $E_{L,\alpha}(\omega)\propto \omega^{-2}$ holds.
Rather, $E_{L,\alpha}(\omega)$ significantly exhibits the non-universal behavior associated with $\bm{f}$~(Fig.~\ref{fig:power_spe}).

To reveal the formation mechanism of $E_{L,\alpha}(\omega)$, we have developed the scale-decomposition analysis method for the Lagrangian velocity based on the bandpass filtering of the velocity field~[Eq.~\eqref{eq:Lagrangian_velocity_decomposition}].
This method enables us to quantify the role of vortices with different scales in the Lagrangian velocity~(Figs.~\ref{fig:lag_bp_vel_time} and \ref{fig:power_spe_bp_validation}).
The power spectral density $E_{L,\alpha}^{(k_c)}(\omega)$ of the bandpass-filtered Lagrangian velocity at $k_f \lesssim k_c\lesssim 0.3\eta^{-1}$ for different $\bm{f}$ and $\mathrm{Re}_\lambda$ collapses on a single function when normalized by the mean energy dissipation rate $\overline{\epsilon}$ and the Lagrangian timescale $\tau_L^{(k_c)}$ at wave number $k_c$~[Fig.~\ref{fig:power_spe_bp}(b)].
Thus, small-scale vortices in the Eulerian inertial range contribute to the Lagrangian velocity in a self-similar way, with a scale-dependent amplitude and timescale consistent with the Kolmogorov similarity hypothesis.
In contrast, the largest-scale flows exhibit non-universal dynamics dominated by the external force rather than the Kolmogorov similarity hypothesis.
Specifically, $E_{L,\alpha}^{(k_c)}(\omega)$ at $k_c\simeq k_f$ deviates from the universal behavior and strongly depends on $\bm{f}$.
Especially for the steady and anisotropic forcing $\bm{f}^{(\mathrm{V})}$, the normalized $E_{L,x}^{(k_c)}(\omega)$ at $k_c=k_f$ takes larger values compared with $E_{L,\alpha}^{(k_c)}(\omega)$ at $k_f \lesssim k_c\lesssim 0.3\eta^{-1}$~[Fig.~\ref{fig:power_spe_bp}(a)], thus contaminating the Kolmogorov scaling formed by the small-scale vortices in the inertial range.
This deviation from the Kolmogorov similarity hypothesis likely arises because the largest-scale vortices have two different timescales related to their swirling motion and persistence.
Consequently, $E_{L,\alpha}(\omega)$ has a broader non-universal range than $E(k)$, which results in the narrow Lagrangian inertial range.
In fact, $\sum_{k_c(>\sqrt{2})} E_{L,\alpha}^{(k_c)}(\omega)$, where the contribution from the forcing wave number, $k_c=\sqrt{2}(\simeq k_f)$, is removed, exhibits a more universal behavior than $E_{L,\alpha}(\omega)$~(Fig.~\ref{fig:power_spe_bp_sum}).

Furthermore, we have also investigated the universality and non-universality of the power spectral density $E_{E}(\omega)$ of the Eulerian velocity at a fixed location to clarify the specific properties of the Lagrangian velocity.
Assuming the relation $k\sim \omega/u^\prime$ based on the sweeping effect by large-scale flows, $E_{E}(\omega)$ almost collapses on $E(k)$, indicating that $E_{E}(\omega)$ contains spatial information on small-scale structures~(Fig.~\ref{fig:all_spe}).
Consequently, in contrast to $E_{L,\alpha}(\omega)$, which reflects the dynamical aspect of turbulence along the Lagrangian trajectory, $E_{E,\alpha}(\omega)$ shows a universal behavior like $E(k)$ due to the dominance of the sweeping effect by the largest-scale flows~(Fig.~\ref{fig:euler_spe}).
Thus, we have concluded that the remarkable non-universality of $E_{L,\alpha}(\omega)$ caused by $\bm{f}$ is specific to the Lagrangian velocity.

In the present study, we have paved the way for relating the dynamics of the Lagrangian velocity to the hierarchy of coherent vortices in turbulence by proposing the scale-decomposition analysis method for the Lagrangian velocity.
One of the most important conclusions is that the largest-scale flows driven by the external force can have a significant influence over a wide frequency range in $E_{L,\alpha}(\omega)$, whereas small-scale vortices belonging to the inertial range exhibit universal dynamics, which is consistent with the Kolmogorov similarity hypothesis.
In other words, the relation between $E_{L,\alpha}(\omega)$ and $E(k)$ is generally nontrivial due to the influential dynamics of largest-scale flows, contrasting with the relation between $E_{E,\alpha}(\omega)$ and $E(k)$~(see Fig.~\ref{fig:all_spe}).
However, there remain some questions to be addressed in future studies.
First, we have yet to quantitatively clarify the universal behavior of the scale-decomposed spectra $E_{L,\alpha}^{(k_c)}(\omega)$ for wave numbers in the inertial range~(Fig.~\ref{fig:power_spe_bp}), including the functional form and its physical explanation.
Second, it is necessary to understand the dynamics of the largest-scale flows depending on the external force, especially for turbulence with a mean flow.
Thus, it is a crucial future study to apply the scale-decomposition analysis method proposed in this study to turbulence with various mean flows.
These future studies will provide deep insight into the dynamics of the Lagrangian velocity associated with coherent vortices of different sizes, which leads to a better understanding of turbulent mixing and a closure theory in the Lagrangian frame.

\begin{acknowledgments}
The present study was supported by JSPS Grants-in-Aid for Scientific Research (20H0206, 21J21061, and 24KJ0109). 
The simulations were conducted under the auspices of the NIFS Collaboration Research Programs (NIFS22KISS010).
\end{acknowledgments}

\end{document}